\documentclass[english,aps,prl,amsmath,amssymb,reprint,superscriptaddress]{revtex4-1}
\usepackage[T1]{fontenc}
\usepackage[latin9]{inputenc}
\usepackage{color}
\usepackage{babel}
\usepackage{array}
\usepackage{textcomp}
\usepackage{multirow}
\usepackage{amsmath}
\usepackage{graphicx}
\usepackage[unicode=true,
 bookmarks=true,bookmarksnumbered=false,bookmarksopen=false,
 breaklinks=false,pdfborder={0 0 0},pdfborderstyle={},backref=false,colorlinks=true,pdfpagemode=FullScreen]
 {hyperref}
\hypersetup{pdftitle={Universal building block for (110)-family surfaces of silicon and germanium},
 pdfauthor={Ruslan Zhachuk and Alexander Shklyaev},
 pdfkeywords={Silicon, Surface, Structure, STM, DFT},
 allcolors=blue}
\usepackage{breakurl}

\makeatletter

\newcommand{\noun}[1]{\textsc{#1}}
\providecommand{\tabularnewline}{\\}

\makeatother

\begin{document}

\title{Universal building block for $(1\,1\,0)$-family silicon and germanium
surfaces}

\author{Ruslan Zhachuk}
\email{zhachuk@gmail.com}

\affiliation{Institute of Semiconductor Physics, pr. Lavrentyeva 13, Novosibirsk
630090, Russia}

\author{Alexander Shklyaev}

\affiliation{Institute of Semiconductor Physics, pr. Lavrentyeva 13, Novosibirsk
630090, Russia}

\affiliation{Novosibirsk State University, ul. Pirogova 1, Novosibirsk 630090,
Russia}

\date{\today}

\begin{abstract}
The universal building block which is an essential part of all atomic
structures on $(1\,1\,0)$ silicon and germanium surfaces and their
vicinals is proposed by combining first-principles calculations and
scanning tunneling microscopy (STM). The atomic models for the $(1\,1\,0)-(16\times2)$,
$(1\,1\,0)-c(8\times10)$, $(1\,1\,0)-(5\times8)$ and $(17\,15\,1)-(2\times1)$
surface reconstructions are developed on the basis of the building
block structure. The models exhibit very low surface energies and
excellent agreements with bias-dependent STM images. It is shown that
the Si$(47\,35\,7)$ surface shares the same building block. Our study
closes the long-debated pentagon structures on $(1\,1\,0)$ silicon
and germanium surfaces.
\end{abstract}
\maketitle
\begin{figure}
\includegraphics[clip,width=8cm]{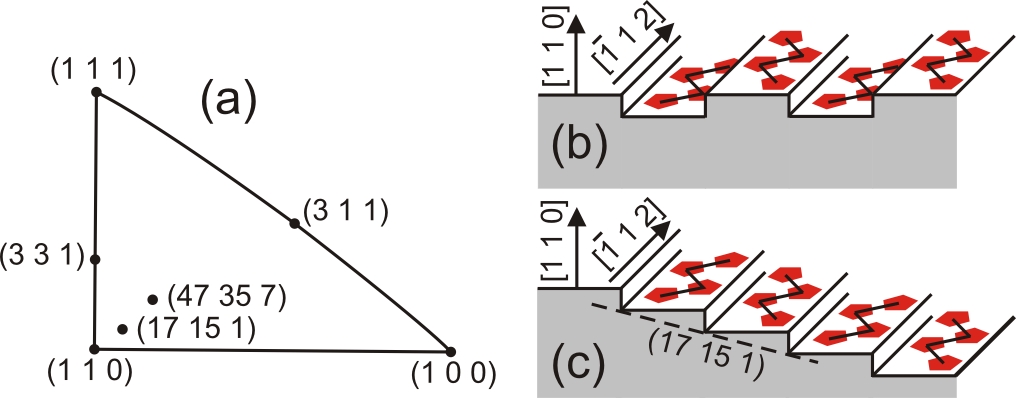}

\caption{\label{fig1} (a) Unit stereographic triangle with some stable silicon
planes marked. (b), (c) Schematic views of $(1\,1\,0)-(16\times2)$
(b) and $(17\,15\,1)-(2\times1)$ (c) silicon and germanium surfaces. }
\end{figure}

Over the past three decades, large efforts have been made to understand
the atomic and electronic structure of $(1\,1\,0)$ silicon and germanium
surfaces using scanning tunneling microscopy (STM) and spectroscopy,
photoelectron spectroscopy and first principles calculations \cite{yam94,pack97,gai98,an00,ste04,ich03,ich04,ich04a,set09,set10,set11,yam16,tak01,mul14,yam00,ohi07,kim07,nag12,loen88,bam14}.
This indicates both the complexity of the task and its high scientific
relevance. One of the reasons for the persistent interest to these
surfaces is their peculiar properties, such as high hole mobility
in the devices fabricated on the Si$(1\,1\,0)$ surface \cite{che06}
and strong surface anisotropy. The second feature became especially
atrractive due to the recent success in the formation of single-domain
$(16\times2)$ reconstruction on the Si$(1\,1\,0)$ surface \cite{yam07,lew17}.
This makes $(1\,1\,0)$ surfaces very convenient substrates for the
growth of one-dimensional objects, such as nanowires \cite{mas06,hong11,hon11,hus13,wat16,hon16}.
It is also worth noting that, among all low-index silicon and germanium
surfaces, $(1\,0\,0)$, $(1\,1\,1)$ and $(1\,1\,0)$ {[}Fig.~\ref{fig1}(a){]},
only the $(1\,1\,0)$ structure is still not understood, and, therefore,
it is of signifcant academic interest as well. 

The common feature of all reconstructed $(1\,1\,0)$ silicon and germanium
surfaces is the presence of bright spots exhibiting pentagonal or
tetragonal shapes (hereafter polygons) in high-resolution STM images
depending on acquisition conditions \cite{an00,set11}. When the Ge$(1\,1\,0)$
surface is observed at an elevated temperature (above $430\,\mathrm{{^\circ}C}$),
the polygons are closely packed and show no long range order \cite{ich95,ich04a,mul14}.
However, when the temperature is lowered to about $380\,\mathrm{{^\circ}C}$
the polygons begin to line up and their density is lowered, indicating
the formation of the $c(8\times10)$ reconstruction. A very long annealing
at $380\,\mathrm{{^\circ}C}$ converts the $c(8\times10)$ reconstruction
into the $(16\times2)$ surface structure. Thus, the $(16\times2)$
reconstruction is equilibrium, while the $c(8\times10)$ surface structure
is only a transient (metastable) structure of the Ge$(1\,1\,0)$ surface.
The structural transformations on the Si$(1\,1\,0)$ surface are similar
to those of the Ge$(1\,1\,0)$ surface, but the transient structure
is $(5\times8)$ \cite{ohi07}. The formation of polygons in varied
experimental conditions at various temperatures is a strong indication
of their exceptional stability owing to low formation energy values.

Perhaps, the most widely known structure of $(1\,1\,0)$ silicon and
germanium surfaces is the structure referred to as $(16\times2)$.
This notation, however, does not follow the Wood\textquoteright s
notation \cite{wood64}, since this surface reconstruction can only
be correctly described by a matrix \cite{ste04a}. According to the
STM data, the $(1\,1\,0)-(16\times2)$ surface is composed of a periodic
up-and-down sequence of terraces with the height difference equal
to a single $(1\,1\,0)$ atomic layer and $\approx25\,\mathrm{\mathring{A}}$
step-step separation {[}Fig.~\ref{fig1}(b){]}. The terraces exhibit
the zig-zag chains of polygons along step edges. Thus, the structure
appears as the equidistant stripes running in $<\bar{1}12>$ directions
and forming a natural substrate for the nanowire growth. Since the
$(16\times2)$ reconstruction is an equilibrium structure of $(1\,1\,0)$
silicon and germanium surfaces, it should provide a minimum free surface
energy. Considering that the $(1\,1\,0)$ terraces in the $(16\times2)$
reconstruction are relatively wide, the step-step interactions should
be small and each step could be considered independently in the first
approximation. Taking this note into account, the reconstructed $(1\,1\,0)$
terraces, linked in a manner of ``down, down, down, ...'' {[}Fig.~\ref{fig1}(c),
or ``up, up, up ...''{]} should also provide a free energy minimum
and, therefore, be observed in experiments. A simple calculation shows
that these surfaces would have $\left\{ 17\,15\,1\right\} $ orientations
{[}$4.4{^\circ}$ off from the $(1\,1\,0)$ plane, Figs.~\ref{fig1}(a)
and (c){]}. Indeed, small facet planes of the $\left\{ 17\,15\,1\right\} $
orientation were found on $(1\,1\,0)$ silicon and germanium surfaces
\cite{olsh77,yam94a,yam94,yam00,ich04a,mul14}. Therefore, there is
a tight structural relationship between the $(17\,15\,1)-(2\times1)$
and $(1\,1\,0)-(16\times2)$ surfaces. Since these surfaces are built
from the same structural elements, they should have very close formation
energies and electronic structures. 

Thirteen full and partial atomic models were proposed to explain the
structures found on $(1\,1\,0)$ silicon and germanium surfaces \cite{nest90,yam94,pack97,gai98,an00,ste04,ich03,ich04,ich04a,set09,set10,set11,yam16}.
While there is a general consensus about the positioning of polygons
in different surface reconstructions, the main difficulty is the atomic
structure of the polygon itself. In this Letter we develop a realistic
model of the building block which appears as polygons in the STM images
of $(1\,1\,0)$ silicon and germanium surfaces and their vicinals.
Using this block we built the microscopic models of the $(16\times2)$,
$c(8\times10)$ and $(5\times8)$ reconstructions of the $(1\,1\,0)$
surfaces which show a remarkably low surface energy and closely reproduce
the experimental bias-dependent STM images. We demonstrate that the
vicinal $(1\,1\,0)$ surfaces, such as $(17\,15\,1)$ and $(47\,35\,7)$,
also share the same universal building block.

The calculations were carried out using the pseudopotential \cite{tro91}
density functional theory \textsc{siesta} code \cite{sol02} within
the local density approximation (LDA) to the exchange and correlation
interactions between electrons \cite{per92}. The valence states were
expressed as linear combinations of the Sankey-Niklewski-type numerical
atomic orbitals \cite{sol02}. In the present calculations, the polarized
double-$\mathrm{\zeta}$ functions were assigned for all species.
This means two sets of\emph{\noun{ }}$s-$ and $p-$orbitals plus
one set of $d-$orbitals on silicon and germanium atoms, and two sets
of $s-$orbitals plus a set of $p-$orbitals on hydrogen atoms. The
electron density and potential terms were calculated on a real space
grid with the spacing equivalent to a plane-wave cut-off of $200\,\mathrm{Ry}$.

Following the other authors \cite{ste04,yam16}, we neglect the contribution
of entropy to the surface free energy and evaluate only the leading
term (surface formation energy). In this work, we calculate the surface
energy gain ($\triangle\gamma$), due to the reconstruction and relaxation,
with respect to the bulk-terminated $(1\,1\,0)$ silicon and germanium
surfaces \cite{zha17a}. $\triangle\gamma$ values were calculated
using 6 layers thick slabs (7 layer slabs for the $(16\times2)$ reconstruction)
terminated by hydrogen from one side. A $18\,\mathrm{\mathring{A}}$
thick vacuum layer was used. We used specific $\mathbf{k}$-point
grids for each surface reconstruction/slab, depending on its respective
lateral dimensions, namely: $2\times2\times1$ for $(1\,1\,0)-(5\times8)$
and $c(8\times10)$, $1\times4\times1$ for $(1\,1\,0)-(16\times2)$,
and $2\times4\times1$ for $(17\,15\,1)-(2\times1)$ \cite{mon76}.
The geometry was optimized until all atomic forces became less than
$0.01\,\mathrm{eV/\mathring{A}}$. For the calculations of silicon
and germanium chemical potentials we used the respective bulk supercells
(with equilibrium lattice constants $a_{Si}=5.420\,\mathrm{\mathring{A}}$
and $a_{Ge}=5.650\,\mathrm{\mathring{A}}$) with the lateral dimensions
and $\mathbf{k}$-point grids identical to those used for slab calculations.
The constant-current STM images were produced based on the Tersoff-Hamann
approximation \cite{ter83} using the eigenvalues and eigenfunctions
of the Kohn-Sham equation \cite{koh65} for a relaxed atomic structure.
The surface optimized basis set (cut-off radii for $s-$, $p-$, and
$d-$orbitals are $R_{s}=9\,\mathrm{Bohr}$, $R_{pd}=11\,\mathrm{Bohr}$
for Si ) was used for STM image calculations \cite{zha18}.

\begin{figure}
\includegraphics[clip,width=8cm]{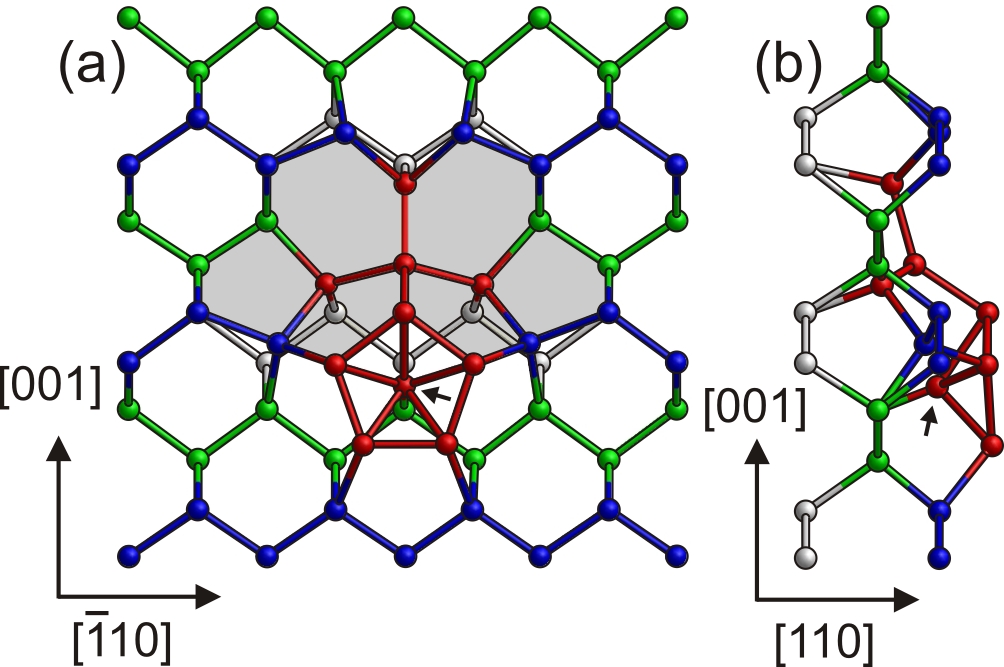}

\caption{\label{fig2} Universal building block (UBB) structure of reconstructed
$(1\,1\,0)$ silicon and germanium surfaces. The zig-zag atom chains
in the $[\bar{1}10]$ direction are the atoms of the first, second
and third layers marked in blue, green and white, respectively. The
additional atoms and atoms strongly shifted from their ideal $(1\,1\,0)$
lattice positions are red. The interstitial atoms are highlighted
by arrows. (a) Top view. The rebonded area is shaded. (b) Side view. }
\end{figure}

The STM images were recorded at room temperature in the constant-current
mode using an electrochemically etched tungsten tip. The measurements
were performed in an ultrahigh vacuum chamber ($7\times10^{-11}\,\mathrm{Torr}$)
on a system equipped with an STM (\textsc{OMICRON}). A clean Si$(47\,35\,7)$
surface was prepared by the sample flash annealing at $1250\,\mathrm{{^\circ}C}$
for $1\,\mathrm{min}$ followed by a stepwise cooling with an average
rate of $\approx5\,\mathrm{{^\circ}C/min}$ within a temperature range
from about 850 to $400\,\mathrm{{^\circ}C}$. The \textsc{WSxM} software
was used to process the experimental and calculated STM images \cite{hor07}. 

The structure of universal building block (UBB), proposed in this
Letter, is shown in Figs.~\ref{fig2}(a) and \ref{fig2}(b). It has
the mirror symmetry with respect to the $(\bar{1\,}1\,0)$ plane.
The UBB consists of the interstitial atom, which holds together five
atoms of the surrounding pentamer, and closely integrated rebonded
area {[}shaded area in Fig.~\ref{fig2}(a){]}. The pentamers with
interstitial atoms were succesfully applied to develop the atomic
models of reconstructed Si$(1\,1\,3)$ and Si$(3\,3\,1)$ surfaces
\cite{dab94,zha17}. Unfortunately, porting this structural unit to
other surfaces is not straightforward, since these surfaces have different
bond configurations. The rebonded area contains four pentagonal rings
and two hexagonal rings {[}shaded area in Fig.~\ref{fig2}(a){]}.
The UBB structure eliminates 8 dangling bonds on the unreconstructed
$(1\,1\,0)$ surface and requires a very little mass transfer to be
built, since each UBB contains only 3 additional atoms.

\begin{figure}
\includegraphics[clip,width=8cm]{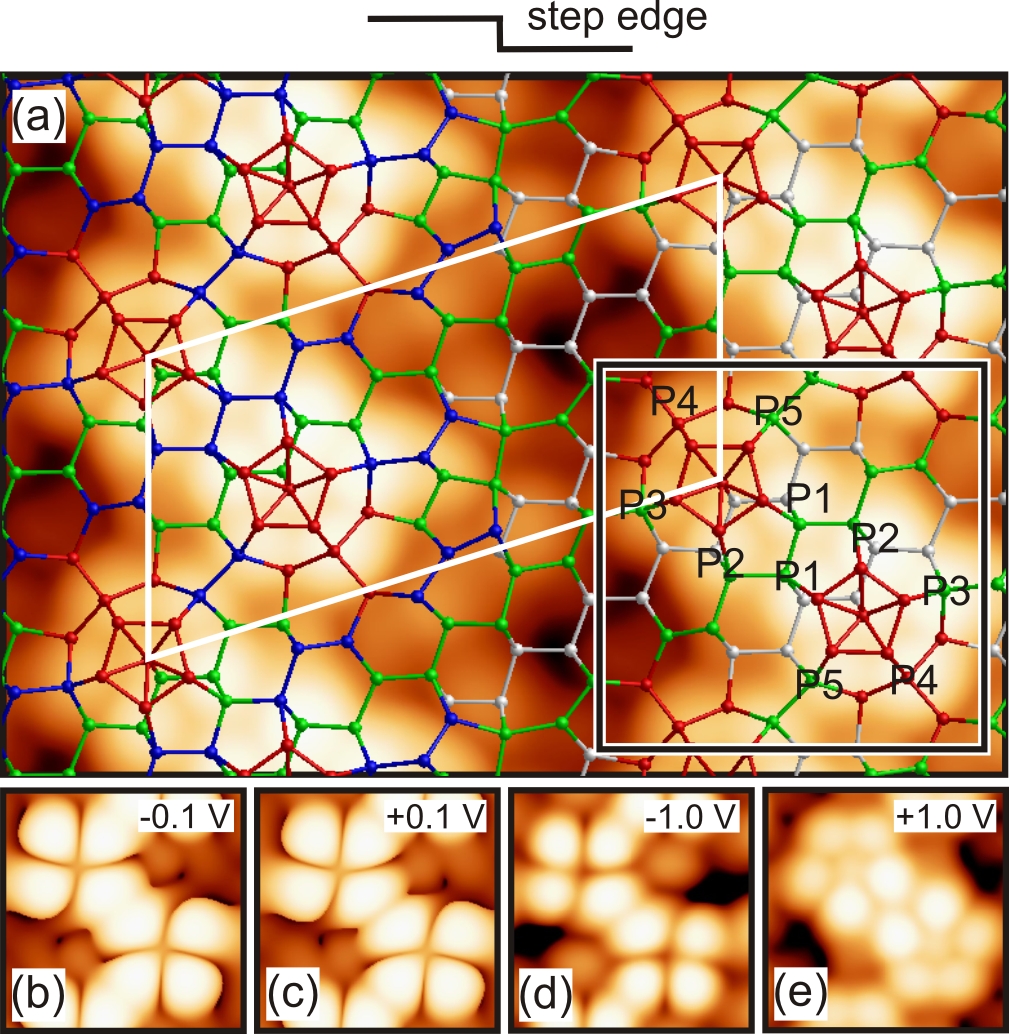}

\caption{\label{fig3} (a) UBB model of the Si$(17\,15\,1)-(2\times1)$ surface
and the corresponding calculated constant-current STM image assuming
$U=+1.0\,\mathrm{V}$. Zig-zag atom chains are the atoms in the $[\bar{1}10]$
direction of the first three $(1\,1\,0)$ layers marked in blue, green
and white. The UBBs atoms are marked in red. The step edge (vertical)
direction is $[\bar{1}12]$. The pentamer pairs are highlighted by
a black square, the unit cell is shown by a white parallelogram. The
pentamer spots are labeled P1-P5 according to Ref.~\onlinecite{set11}.
(b)-(e) The STM images of pentamer pairs on the Si$(17\,15\,1)-(2\times1)$
surface calculated assuming different applied voltages. }
\end{figure}

In Fig.~\ref{fig3}(a) are the UBB model of the $(17\,15\,1)-(2\times1)$
surface and the corresponding calculated constant-current STM image.
The pentamer spots are labeled P1-P5 according to Ref.~\onlinecite{set11}.
The model illustrates both the arrangement of pentamers on $(1\,1\,0)$
terraces and the step edge structure on $(17\,15\,1)-(2\times1)$
and $(1\,1\,0)-(16\times2)$ surfaces. The step edge structure shown
in Fig.~\ref{fig3}(a) is different from the structure proposed in
Ref.~\onlinecite{set10}. See Supplemental Material {[}\emph{URL
will be inserted by publisher}{]} for UBB models and the corresponding
calculated STM images of $(16\times2)$, $c(8\times10)$ and $(5\times8)$
reconstructions on the $(1\,1\,0)$ surface. We note here that the
UBB has a 3D structure, which is only partially accessible for STM
observations. The interstitial atoms, as well as the rebonded area
atoms, are difficult to visualize, since all their bonds are saturated
and they are located somewhat below the surface level. The visible
pentamer size in the calculated STM image in Fig.~\ref{fig3}(a)
is larger than that from the atomistic model. The same statement holds
for the experimental STM images. This size mismatch sustained an argument
against the atomic models based on pentamers with interstitial atoms
in the past \cite{sak09}. The seeming disagreement, however, is explained
by the presence of polarized surface radical states showing a pronounced
angle with respect to the surface normal \cite{zha17b,zha18,zha19}. 

The constant-current STM images of pentamer pairs, outlined by the
black square in Fig.~\ref{fig3}(a), were calculated for the empty/filled
electronic states and for low/moderate applied voltage to compare
them with the available experimental data. The pentamers exhibit four
lobes when observed using filled electronic states and five lobes
when empty electronic states are used at the $1.0\,\mathrm{V}$ applied
bias {[}Figs.~\ref{fig3}(d) and (e){]} in agreement with the experimental
data \cite{an00,set11}. At the low bias ($0.1\,\mathrm{V}$), however,
the pentamers show only four lobes at both polarities {[}Figs.~\ref{fig3}(b)
and (c){]}. Again, this is in a full agreement with the experimental
STM images reported in Ref.~\onlinecite{set11}. The detailed inspection
of the STM image in Fig.~\ref{fig3}(e) reveals that the spots P1
and P2 facing the neighboring pentamer are brighter than other spots
(P3-P5). Splitting the four-lobe pattern into five-lobes at the positive
bias and increased intensity of P1 and P2 spots is caused by the empty
state superposition at about $0.5\,\mathrm{eV}$, as it was experimentally
demonstrated by Setv{\'i}n \emph{et al.} \cite{set11}. Note that
both shape and relative intensities of electronic states are very
sensitive to the atomic structure and the UBB model reproduces them
very well, while other models fail to describe even basic pentamers
geometry. This is a strong argument for the validity of UBB model.
Another argument is very low formation energy values for all surface
structures composed of UBBs, as shown below.

\begin{table}
\begin{ruledtabular}
\caption{\label{tab1}Reconstruction-induced energy gain $(\mathrm{\triangle\gamma,\,meV/\mathring{A}}^{2})$
for different models of the Si$(1\,1\,0)-(16\times2)$ reconstruction
with respect to the bulk-terminated Si$(1\,1\,0)$ surface: ATI \cite{ste04},
THTR stepped \cite{yam16} and UBB.}
\begin{tabular}{cc}
Model & $\triangle\gamma$ \tabularnewline
\hline 
ATI  & 21.4, 21.6 \cite{yam16}, 23.8 \cite{ste04}\tabularnewline
\multirow{2}{*}{THTR stepped } & \multirow{2}{*}{ 32.0, 30.5 \cite{yam16}}\tabularnewline
 & \tabularnewline
UBB & 31.5 \tabularnewline
\end{tabular}
\end{ruledtabular}

\end{table}

Since many atomic models of the $(1\,1\,0)$ surface structures have
been proposed, we will compare our $(1\,1\,0)-(16\times2)$ UBB model
only with two other models: the adatom-tetramer-interstitial (ATI)
model, the most cited in the literature \cite{ste04} and the recently
proposed tetramer heptagonal- and tetragonal-ring (THTR) stepped structure
\cite{yam16}. In Table~\ref{tab1} are the energy gains for three
different $(16\times2)$ reconstruction models. The ATI model by Stekolnikov
\emph{et al. }can be ruled out since it shows the low energy gain.
In addition, the respective calculated constant-current STM images
do not match the experimental STM images of Si$(1\,1\,0)-(16\times2)$,
although the ATI model also contains pentamers with interstitial atoms.
According to our calculations, the recently proposed THTR stepped
model is slightly favored over UBB, when using LDA to the exchange
and correlation interactions {[}Tab.~\ref{tab1}{]} \cite{per92},
but the opposite trend is observed when the generalized gradient approximation
(GGA) \cite{per97} is used. Anyway, the energy difference between
THTR stepped and UBB models is within the typical error of about $1\,\mathrm{meV/\mathring{A}^{2}}$
for these type of calculations. Finally, we cannot exclude that the
step edges on the $(1\,1\,0)-(16\times2)$ surface have another atomic
configuration than the one shown in Fig.~\ref{fig3}(a). The THTR
stepped model has a serious flaw, since the static model is incompatible
with the experimental STM images of pentamers \cite{yam16}. It was
suggested that the model can reproduce STM images only when the dynamic
buckling of reconstruction elements at room temperature is considered.
There are two objections for this hypothesis. First, there is no indication
that the pentamers at $T=78\,\mathrm{\text{\textdegree}K}$ look different
than that at room temperature in the low-temperature STM study by
Setv{\'i}n \emph{et al.} \cite{set11}. In fact, the pentamers look
basically the same both in low- and in room-temperature STM images
\cite{set00}. Second, the calculated constant-current STM images,
averaged using two buckled surface configurations, do not reproduce
the experimental STM images of pentamers as well (see Supplemental
Material {[}\emph{URL will be inserted by publisher}{]} for the calculated
STM images of the ATI and THTR stepped models of Si$(1\,1\,0)-(16\times2)$).

\begin{table}
\begin{ruledtabular}
\caption{\label{tab2}Number of UBBs per $(1\,1\,0)-(1\times1)$ cell $(n$),
excess coverage ($\triangle\varTheta$, monolayers) and reconstruction-induced
energy gain $(\mathrm{\triangle\gamma,\,meV/\mathring{A}}^{2})$ with
respect to the bulk-terminated $(1\,1\,0)$ surface for various UBB-based
structural models. }
\begin{tabular}{ccccc}
\multirow{2}{*}{Reconstruction} & \multirow{2}{*}{$n$ } & \multirow{2}{*}{$\triangle\varTheta$} & \multicolumn{2}{c}{$\triangle\gamma$ }\tabularnewline
\cline{4-5} 
 &  &  & \multirow{1}{*}{Si} & \multirow{1}{*}{Ge}\tabularnewline
\hline 
$(16\times2)$ & 0.125 & 0.66 & 31.5 & 28.2\tabularnewline
\multirow{1}{*}{$(5\times8)$} & 0.1  & 0.15  & \multirow{1}{*}{30.4} & \multirow{1}{*}{28.1}\tabularnewline
$c(8\times10)$ & 0.1  & 0.15  & 30.2 & 28.1\tabularnewline
\end{tabular}
\end{ruledtabular}

\end{table}

In table~\ref{tab2} is the number of UBBs per $(1\,1\,0)-(1\times1)$
cell, excess coverage and reconstruction-induced energy gain with
respect to the bulk-terminated $(1\,1\,0)$ surface for various UBB-based
structural models. All atomic structures show noticeably large energy
gains, while the gain for $(16\times2)$ structure is slightly higher
than that for other two reconstructions. This is in agreement with
the experimental results indicating that this structure is equilibrium
\cite{ich95,ich04a,mul14,ohi07}. There is a correlation between the
highest energy gain and the highest surface density of UBBs in the
$(16\times2)$ structure {[}Tab.~\ref{tab2}{]}. This finding can
be interpreted in the way that the main physical reason for the low
energy of reconstructed $(1\,1\,0)$ surfaces is the presence of UBBs.
Further, the formation of $(5\times8)$ and $c(8\times10)$ structures
requires a very little mass transfer (addition of 0.15 monolayers),
since it solely depends on the UBBs assembly. The $(16\times2)$ structure
requires a larger mass transfer (addition of 0.66 or removal of 0.34
monolayers), since, in addition, it involves the formation of up and
down steps. This observation clarifies the nature of metastable $(5\times8)$
and $c(8\times10)$ structures; namely, it shows that these structures
are formed due to a limited mobility of surface atoms. A similar formation
of metastable surface structures after rapid sample cooling was observed
on the Si$(1\,1\,1)$ surface \cite{zhao98,yang94}.

\begin{figure}
\includegraphics[width=8cm]{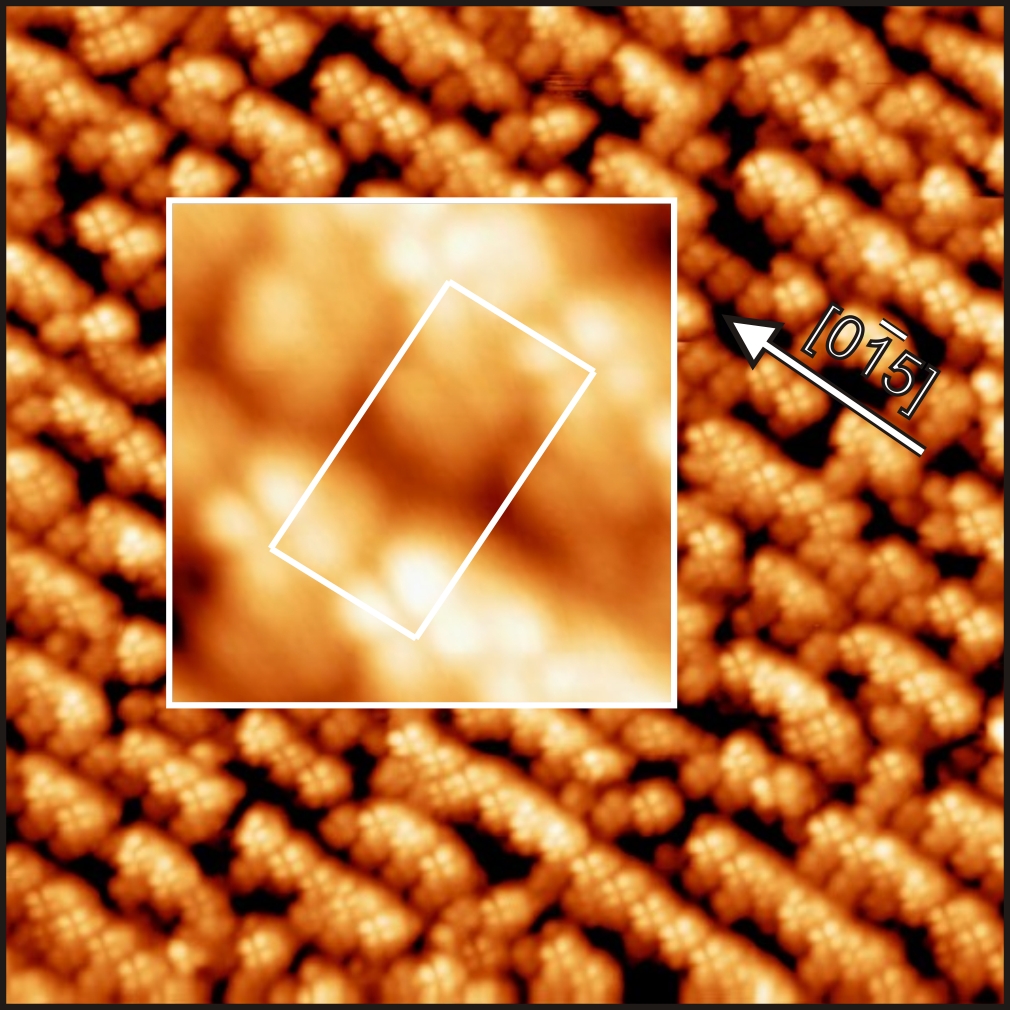}

\caption{\label{fig4}Experimental STM image of the Si$(47\,35\,7)$ surface,
$400\mathrm{\times400\,\mathring{A}}^{2}$. $U=-0.8\,\mathrm{V}$,
$I=2.0\,\mathrm{pA}$. In the inset is a high-resolution STM image
of the structure with the unit cell outlined, $5\mathrm{\times5\,\mathring{A}}^{2}$. }
\end{figure}

Finally, in Fig.~\ref{fig4}, we show the experimental STM image
of the stable Si$(47\,35\,7)$ surface. This vicinal Si$(1\,1\,0)$
surface, inclined at about $10.7{^\circ}$ degrees to the $(1\,1\,0)$
plane {[}Fig.~\ref{fig1}(a){]} was first observed and identified
by Olshanetsky and Shklyaev using low-energy electron diffraction
\cite{olsh79}. The STM image reveals the ordered reconstructed surface
with the structural blocks aligned into straight chains along the
$[0\bar{1}5]$ direction. The high resolution STM image shown in the
inset clearly exhibits the four-lobe patern for each structural block
at the unit cell corners. The pattern is specific for the pentamers
observed at negative bias on $(1\,1\,0)$ silicon and germanium surfaces
{[}Fig.~\ref{fig3}(d){]}. Since the $(1\,1\,0)$ terraces on the
Si$(47\,35\,7)$ surface can accomodate a single UBB in width, we
assume that this surface is related to the same family of reconstructed
surfaces as $(1\,1\,0)$ and $(17\,15\,1)$.

In summary, we have shown that all reconstructions found on $(1\,1\,0)$
silicon and germanium surfaces and their vicinals share the same building
block. The atomic structure of the universal building block is proposed
and it has been demonstrated that the surfaces composed of these blocks
possess very low surface energies and show excellent agreement with
the bias-dependent experimental STM images. Our study concludes the
long-standing debate on the atomic structures of $(1\,1\,0)$ silicon
and germanium surfaces, consistently describing their reconstructions
on a single basis.
\begin{acknowledgments}
We would like to thank the Novosibirsk State University for providing
the computational resources. This work was supported by the Russian
Foundation for Basic Research (Project No. 18-02-0000025).
\end{acknowledgments}

\bibliographystyle{apsrev4-1}

\end{document}